\documentclass{ws-ijmssc}
	
\usepackage{epsfig}

\begin{document}

\markboth{H. Jiang et al}{A New Metric For Routing In Military Wireless Network}

%%%%%%%%%%%%%%%%%%% Publisher's Area please ignore %%%%%%%%%%%%%%%%%%%%%%%
\catchline{}{}{}{}{}
%%%%%%%%%%%%%%%%%%%%%%%%%%%%%%%%%%%%%%%%%%%%%%%%%%%%%%%%%%%%%%%%%%%%%%%%%%

\title{A NEW METRIC FOR ROUTING IN MILITARY WIRELESS NETWORK\ }

\author{HAIBO JIANG ,YAOFEI MA ,DONGSHENG HONG, ZHEN LI\ }

\address{School of Automation Science and Electrical Engineering \\ Beihang University, No.37 Xue Yuan Road\\
Hai Dian District, Beijing 100191, China\\\
jianghaibohai\_jin@126.com}
\maketitle

\begin{history}
\received{(Day Month Year)}
%\revised{(Day Month Year)}
\accepted{(Day Month Year)}
%\comby{(xxxxxxxxxx)}
\end{history}

\begin{abstract}
Wireless Ad-hoc network is generally employed in military and emergencies due to its flexibility and easy-to-use. It¡¯s suitable for military wireless network that has the characteristics of mobility and works effectively under severe environment and electromagnetic interfering conditions. However, military network cannot benefit from existing routing protocol directly; there exists quite many features which are only typical for military network. This paper presents a new metric for routing, which is employed in A* algorithm. The goal of the metric is to choose a route of less distance and less transmission delay between a source and a destination. Our metric is a function of the distance between the ends and the bandwidth over the link. Moreover, we take frequency selection into account since a node can work on multi-frequencies. This paper proposed the new metric, and experimented it based on A* algorithm. The simulation results show that this metric can find the optimal route which has less transmission delay compared to the shortest path routing.
\end{abstract}


\section{Introduction}	

In recent years, massive efforts have been directed towards mobile ad-hoc network. In this kind of network, for there is no ready infrastructure, when two indirect nodes need to transmit information between each other, they depend on the relay of other nodes\cite{1}. How to find a route from source node to the destination node is a key problem. In general, AODV, DSR and TORA routing protocols are commonly used\cite{2}. However, most routing protocols find their route by the shortest path criterion, which is not enough for battlefield ad-hoc network. This is illustrated by the following reasons. Firstly, in order to reduce the transmission delay, it is indispensable to incorporate the bandwidth into the new metric, which can improve the network capacity and accuracy of reconnaissance. Secondly, most existing protocols do not consider various frequency that communication devices own so they are not suitable for such situation where nodes equipped with many communication devices with different frequencies, which is very common in military wireless network. Military forces usually include vehicles that are equipped with several radios that each works at different frequency. Radios that have same frequency can communicate. As a result, searching for the communication route should take frequency-matching into account.

Finally, if two vehicles could not communicate directly, they have to rely on the relay of other vehicles, leading to internal communication and external communication since a vehicle can have more than one radio in the network. Figure 1 depicts the issue.
\begin{figure}[th]
\centerline{\psfig{file=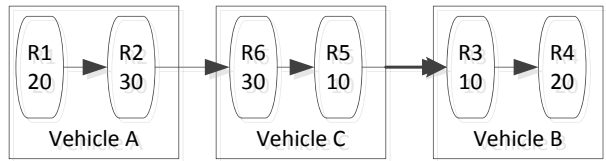,width=8cm}}
\vspace*{8pt}
\caption{The forwarding process among vehicles with multi-radios.}
\end{figure}
There are three vehicles in Figure1. Each has two radios respectively. The operating frequency is given in Figure 1. If R1 wants to communicate with R4, it has to connect with vehicle C firstly, and then reach to vehicle B for frequency matching. R1 and R2 reside on the same vehicle. They can communicate directly through internal path. R2 have the same operating frequency with R6, so R2 can connect to R6. Similarly, R6 can say to R5 directly, R5 connects to R3, and R3 reaches to R4. Our routing algorithm will consider the internal and external path between radios.

The objective of the metric proposed is to construct robust, reliable and less-hop network that is closer to real world scenario. We employ A* algorithm to solve this problem. As a classical heuristic search algorithm, A* algorithm is predominant in solving the problem of the path searching \cite{3,4,5,6,7}. A* is extended to achieve three goals. Firstly, when searching the optimal path between source and end nodes, the algorithm will choose the next hop with the minimum routing metric. Secondly, A* resembles reactive routing protocol since it is computed once communication happens. So the node needn¡¯t maintain the topology of the network. Hence, it will decrease the routing overhead significantly. Finally, A* does frequency-selection work and addresses the internal communication issue aforementioned elegantly.

The structure of this paper is organized as follows. In Section 2, we give a brief introduction to A* algorithm and then propose our new metric. In section 3, we conduct a simulation experiment to find the optimal path following the proposed metric in a military wireless network scenario. Finally, we conclude with future work.

\section{Description Of Metric And The Implementation}

\subsection{A* algorithm }
A* algorithm is an optimization algorithm for heuristic search, with which we can get the optimal solutions in every step of the search\cite{7}. The search is a process to search and find the goal node from the source node by applying evaluation function to sort the nodes\cite{3}. It chooses and maintains the nodes by an OPEN table and a CLOSED table based on an evaluation function defined as:
\begin{equation}
f(n) =g(n)+h(n)
\end{equation}

Where,$g(n)$ denotes the actual routing-cost from the start node to the current node; heuristic function $h(n)$ denotes the estimation of the cost of optimal path from the current node to target node, $h(n)$ depends on the heuristic information of the problem domain. The search of the algorithm may be described as follows:

(1) Put the source node $s_0$ , $f(s_0)$ -attached, into the OPEN table;

(2) If the OPEN table is empty, exit, and the search fails;

(3) Remove node $N_0$  from the OPEN table, whose $f(\cdot)$ should be the smallest in the table. Put node $N_0$  into the CLOSED table, the number of the nodes is n;

(4) If node $N_0$  is the goal node, the search is done, exit.

(5) If node $N_0$ can¡¯t spread, turn to step (2);

(6) Spread node $N_0$ , there will be a group of children nodes, all of which are $f(n)$ -attached. Put these children nodes into the OPEN table, then turn to step (3). For children nodes in this step, some additional steps, as follows, will be done.

(a) Examine the OPEN table and the CLOSED table to find whether the nodes have been included in them. For the nodes that have been included, if they are in the CLOSED table, neglect them; if they are in the OPEN table, it needs to be checked to see whether the value of  $f(n)$ has changed.  $f(n)$ will change following the rule that ¡°choosing the shortest route¡±.

(b) If the nodes have not been included in the OPEN table and CLOSED table, put them into the OPEN table after assigning the back pointer that points to node $N_0$ . And, then sort nodes in OPEN table in ascending order based on $f(n)$ .

The most important part in A* algorithm is to choose proper evaluation function, which is the metric to guide searching process towards the most promising direction for the optimal path.

\subsection{The new routing metric}
In battlefield network, better situation awareness and communication in combat situations will result in higher combat effectiveness. This requires high-bandwidth communications.  Hence, when searching the route, it is necessary to consider the bandwidth of the radio. And less distance guarantees more strength when the radio wave arrives at receiver to gain larger SNR to resist interference.

\subsubsection{The metric design}
Firstly, we list the assumptions about the networks in which the new metric is supposed to operate.
\begin{itemlist}
 \item The size of a data packet transmitted in the network is identical,
 \item All nodes have the ability to forward the messages and they all move in a specific area,
 \item Every node knows other nodes¡¯ positions when searching the path.
\end{itemlist}

Taking above assumptions, a new metric to evaluate the communication route is proposed. The metric¡¯s goal is to choose route with the balance between the length of route and less transmission delay. The metric takes following issues into account:
\begin{itemlist}
 \item the total distance between communication ends. The distance is defined as hops,
 \item the bandwidth of the total route. Sufficient bandwidth will ensure the transmission speed of data.
\end{itemlist}

Let $p$ denotes the metric. Since we tend to select route with shorter distance and larger average bandwidth, it is suitable to let them be inversely proportional.
\begin{equation}
p=\frac{\sum^ n_{i=1} d_i+d_{esti}}{\sum^ n_{i=1}bw_i},i=1...n
\end{equation}

Where,  $d_i$ is the Euclidean distance between two neighboring nodes, $d_{esti}$  denotes the estimation distance from current searching node to destination node,  $bw_i$ is the bandwidth of a node. $n$ is the number of nodes included in current optimal route. The routing protocol selects the route with minimum cost.

\subsubsection{The route searching based on A* algorithm}
If the environment noise is not considered, distance is the only factor determining whether or not the communication link can be built up. If vehicles¡¯ distance is less than the radios¡¯ communication range, the vehicles can link with each other effectively. In the battlefield network,  $g(n)$ is defined as the actual routing metric from the start vehicle to vehicle $N_0$ (assuming $N_0$  is just the vehicle which is being searched.); $h(n)$  denotes the estimated routing metric from vehicle $N_0$ to the goal vehicle.

Before starting route searching process, we first identify the topology of the network by obtaining the neighbors of each node in battlefield and put them into the node¡¯s neighbor list. A node¡¯s neighbors are defined as nodes which are one hop away and have the overlap frequency with the node.

In the operation steps of A* algorithm, we consider multi-frequency selection issue. When spreading vehicle $N_0$ , only vehicles have same frequency as the last vehicle, can be put into the OPEN table. Therefore, it is necessary to traverse all equipped radios on the vehicle. To make it work, every vehicle is treated as a node in the A* algorithm, while the radios of each are attached with it.

Before starting the algorithm, we have to calculate the metric to every other vehicle for every vehicle in the battlefield, and put the vehicles within the minimum path metric into the list of the vehicle. There are two key points when executing the algorithm.
\begin{itemlist}
 \item Traverse every radio on neighbor vehicle and check the frequency,
 \item Check if the neighbor node has one radio at least owing the same frequency as node $N$. If it does, putting the neighbor vehicle into the OPEN table; if it doesn¡¯t, putting the neighbor vehicle into the CLSOED table.
\end{itemlist}

The pseudo code is shown in Figure 2.
\begin{figure}[th]
\centerline{\psfig{file=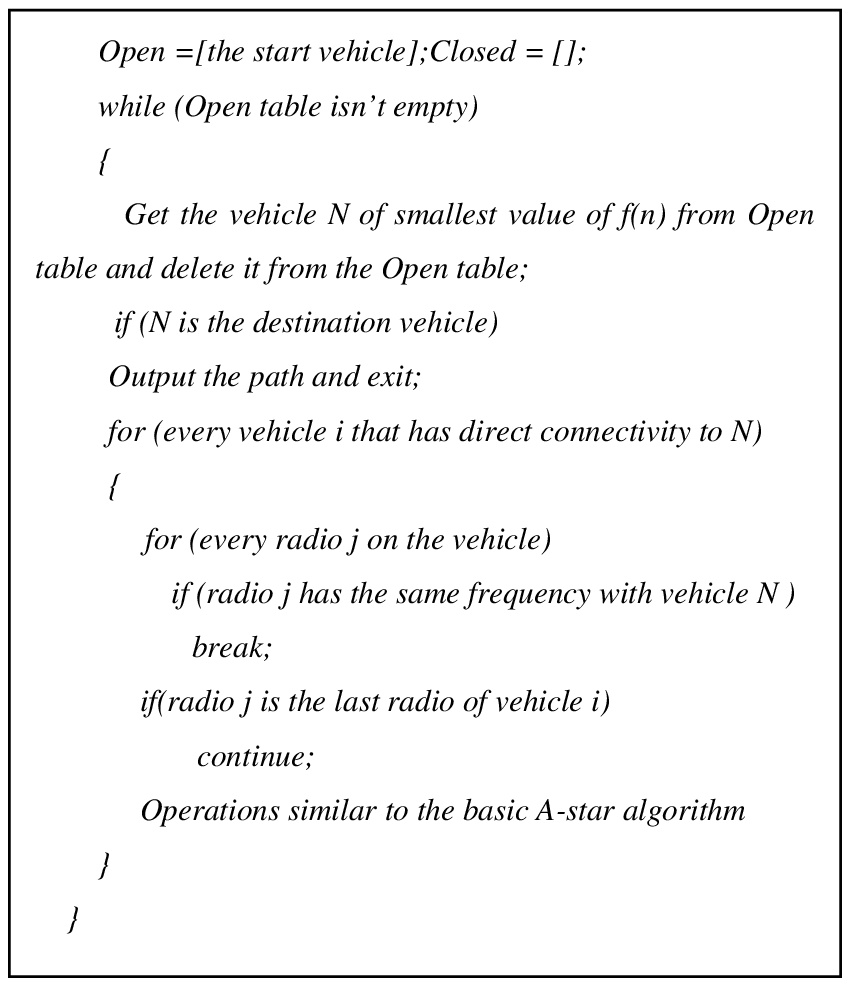,width=8cm}}
\vspace*{8pt}
\caption{Pseudo code of the algorithm.}
\end{figure}

\section{Experiment And Result}
The experiment scenario is designed as follows: 30 vehicles are initially randomly distributed in a square area of $1000m\times1000m$. Two vehicles can communicate directly only if their Euclid distance is less than or equal to the communication threshold distance which is assumed to be 200$m$. The experiments are conducted using the Computer Generated Force $(CGF)$ Software developed by our lab\cite{8,9}.

Firstly, all the vehicles are supposed to have identical frequency to test the validation of the A* algorithm and our new metric. Secondly, vehicles equipped with multi-radios operating on various frequencies are configured to test frequency selection mechanism of our algorithm. Under such situations, we compare the shortest path metric and the new metric to the network which considers both distance and bandwidth factors.

\subsection{Radios with same frequency}
Figure 3 shows the optimal route founded by A* algorithm. The start node is labeled as 1, the destination node is labeled as 7. Because all vehicles have same frequency and same bandwidth, the optimal route is also the shortest route.

The rough solid route is the optimal path found by A* algorithm. The other available routes from node 1 to node 7 are listed in Table 1. They have more hops than the optimal one.

\begin{table}[ht]
\tbl{Several typical routes from node1 to node7.}
{\begin{tabular}{@{}cccc@{}} \toprule
Sequence & Typical path & Sum of path cost\\
& & (Unit: m) \\ \colrule
1\hphantom{00} & \hphantom{00}$1\rightarrow6\rightarrow10\rightarrow7$ & \hphantom{0}444.32  \\
2\hphantom{00} & \hphantom{0}$1\rightarrow6\rightarrow10\rightarrow16\rightarrow7$ & \hphantom{0}448.78  \\
3\hphantom{00} & $1\rightarrow6\rightarrow9\rightarrow14\rightarrow16\rightarrow7$ & \hphantom{0}488.92  \\
4\hphantom{00} & \hphantom{0}$1\rightarrow6\rightarrow9\rightarrow16\rightarrow7$ & \hphantom{0}482.86 \\ \botrule
\end{tabular}}
\end{table}

\begin{figure}[th]
\centerline{\psfig{file=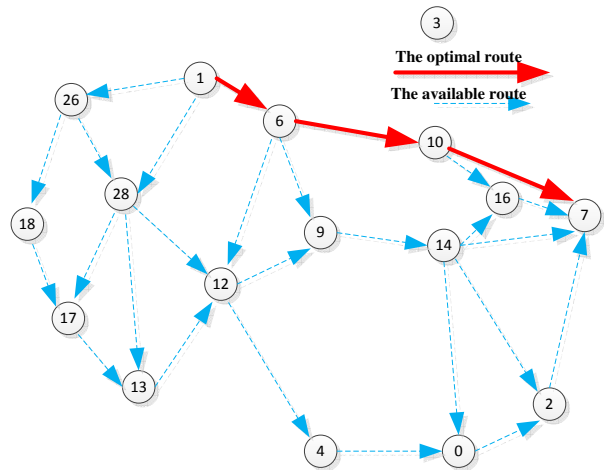,width=8cm}}
\vspace*{8pt}
\caption{The optimal path when all the vehicles have the same frequency.}
\end{figure}

Then, we apply new metric to the experiment. Every radio in the scenario is configured with different bandwidth. The top left labeled number with underline near each node denotes the bandwidth $(Kb/sec)$ of every radio. Figure 4 depicts the result. Since node 28 and node 12 have larger bandwidth than node 6 but are not longer than node 17 and node 13, the algorithm selects node 28 and node 12 as the forwarding node. The result demonstrates a balance between distance and bandwidth. Compared to figure 3, the optimal route in figure 4 may not be the shortest but average bandwidth is larger which contributes to less transmission delay. The average bandwidth operated by the shortest distance and new metric is 5 and 7.4.
\begin{figure}[th]
\centerline{\psfig{file=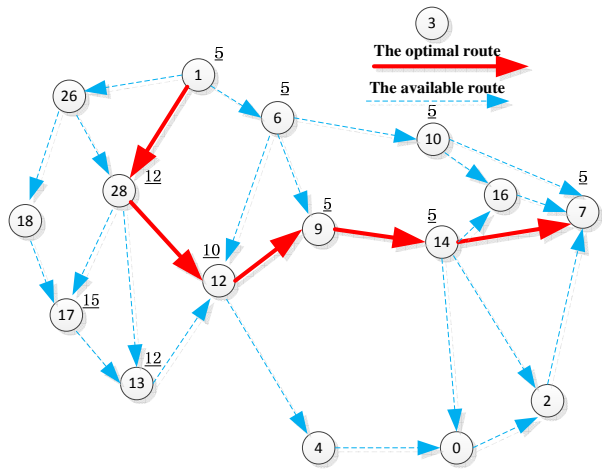,width=8cm}}
\vspace*{8pt}
\caption{The optimal path when vehicles have the different bandwidth.}
\end{figure}

After multi-rounds simulations, the average bandwidth for shortest path and new metric is plotted in Figure 5.
\begin{figure}[th]
\centerline{\psfig{file=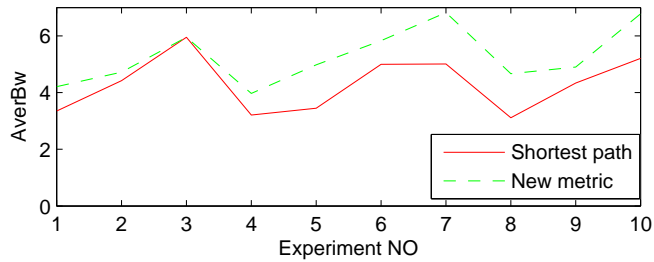,width=10cm}}
\vspace*{8pt}
\caption{Comparison of bandwidth using shortest metric and new metric.}
\end{figure}

Figure 5 shows that the optimal route (the  dotted curve) following the new metric has larger bandwidth than route of shortest route metric (the solid curve). Sometimes the average bandwidth is same due to the shortest route is just the route with larger bandwidth.

\subsection{Radios with different frequencies}
When we consider the multi-radios operation on different frequencies, the optimal route may be not the least hops route. Vehicles with different frequency should be abandoned even their distance is shorter. As Figure 6 shows, the frequency selection will dramatically change the route from source node to destination node.

\begin{figure}[th]
\centerline{\psfig{file=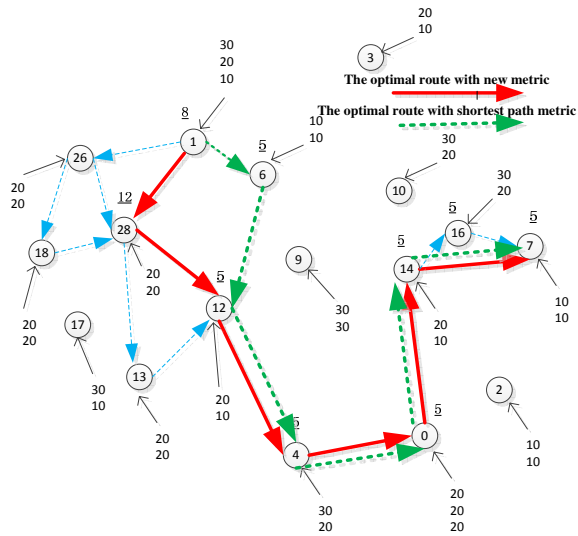,width=8cm}}
\vspace*{8pt}
\caption{The optimal path when vehicles have the different bandwidth.}
\end{figure}
In Figure 6, the right top labeled number near each node indicates the operating frequency of radios equipped on this vehicle. The route with the dotted rough arrow is selected with the shortest path metric under frequency-matching operation.
When the vehicle has the same frequency as the last hop vehicle, it can forward the message, otherwise can¡¯t. Even the distance between node 6 and node 9 is smaller than that between node 6 and node 12, they cannot communicate since they do not have the same frequency. Node 6 to node 12 is the shortest route among all frequency-matching candidates. The result shows frequency matching restriction does affect the search process of the route.

The final experiment considers our new route evaluation metric. Supposing radios reside on a same vehicle operating on same bandwidth. Each data packet is 1024 bytes. The route in figure 6 with rough solid arrow shows the optimal route produced by our new metric.
The top left labeled number with underline near each node denotes the bandwidth $(Kb/sec)$ of every radio. Node 1 chooses node 28 as the last hop node since node 28 has a larger bandwidth than node 6. The average bandwidth is 6 and 6.5 for the green and red route in figure 6 respectively.

\section{Conclusions}
We proposed a new routing metric for A* algorithm, which enables us to find the route with less transmission delay and less distance. Our improved metric are able to search the optimal path when a vehicle can have multiple radios and each radio operates on different frequency. The experiment results show that our new metric outperforms the shortest-path metric considering the time delay.

In future work, we will focuses on high performance routing and re-routing algorithms considering more impact factors, including environment and electromagnetic interference.

\end{document}